\documentclass{ws-procs11x85}
\usepackage{multicol}

\makeindex
\begin{document}
\title{Constraint Algebra in Causal Loop Quantum Gravity}
\author{Fatimah Shojai}
\address{Physics  Department, Tehran University, Tehran, Iran.\\Institute for Studies in Theoretical  Physics and Mathematics,  P.O.Box  19395-5531,  Tehran, Iran.\\E-mail: fatimah@theory.ipm.ac.ir}
\author{Ali Shojai}
\address{Physics  Department, Tehran University, Tehran, Iran.\\Institute for Studies in Theoretical  Physics and Mathematics,  P.O.Box  19395-5531,  Tehran, Iran.\\E-mail: shojai@theory.ipm.ac.ir}
\maketitle\abstract{de-Broglie--Bohm causal interpretation of canonical quantum gravity in terms of Ashtekar new variables is built. The Poisson brackets of (deBroglie--Bohm) constraints are derived and it is shown that the Poisson bracket of Hamiltonian with itself would change with respect to its classical counterpart.}
\begin{multicols}{2}
\baselineskip=13.07pt
\section{Introduction}
Currently\cite{CONSTRAINT} it is shown that using de-Broglie--Bohm causal interpretation of quantum mechanics\cite{BOHM}, one can derive meaningful relations for constraint algebra and the equations of motion. This is done using the \textit{old} variables, i.e. the dynamical variable is chosen to be the metric on spatial slices in an ADM 3+1 decomposition. The new algebra is a clear projection of general coordinate transformation into the spatial and temporal diffeomorphisms.

In ref. \cite{CONSTRAINT} it is shown that the diffeomorphism subalgebra does not change with respect to the classical one. The Poisson bracket of the quantum Hamiltonian and the diffeomorphism constraints which represents the fact that the quantum Hamiltonian is a psedu scalar under diffeomorphisms, is also the same as in the classical case.
Finally the Poisson bracket of the quantum Hamiltonian constraint with itself differs dramatically with its classical counterpart. In fact this Poisson bracket would be zero weakly, i.e. using the equations of motion. This result is just what one expects for the Hamiltonian.

The quantum Hamiltonian is the classical one added with the quantum potential and gives the system quantum trajectories. In the de-Broglie--Bohm interpretation of quantum mechanics one deals with trajectories and thus with Poisson brackets (not commutators). The system has a well--defined trajectory in the phase space obtained from quantum Hamilton--Jacobi equation.
Thus the de-Broglie--Bohm quantum mechanics represents a deterministic picture of particle trajectory consistent with the statistical predictions of the standard quantum mechanics. In the gravitational case, the dynamics of the metric is determined as a modification of classical Einstein's equation by the quantum potential and quantum force. These are covariant under spatial and temporal diffeomorphisms.\cite{CONSTRAINT}

It is proved that the \textit{new} variables of gravity\cite{ASHTEKAR} are more useful in making quantum gravity. So a natural question is how looks like the causal interpretation of canonical quantum gravity in terms of new variables? Also one can ask about the constraints algebra and equations of motion. In this paper we shall answer to these questions.
\section{Causal interpretation in terms of new variables}
In terms of new variables, gravity consists of three constraints, gauge, diffeomorphism and Hamiltonian constraints. The dynamical variables are the self--dual connection $A_a^i$ and the canonical momenta are $\widetilde{E}^a_i$. The constraints are given by:
\begin{equation}
G_i={\cal D}_a \widetilde{E}^a_i
\end{equation}
\begin{equation}
C_b=\widetilde{E}^a_i F^i_{ab}
\end{equation}
\begin{equation}
H=\epsilon^{ij}_k \widetilde{E}^a_i \widetilde{E}^b_j {\cal F}^k_{ab}
\end{equation}
in which ${\cal D}_a$ represents self--dual covariant derivative and $F^i_{ab}$ is self--dual curvature.

Canonical quantization of these constraints can be achieved in the connection representation via changing $\widetilde{E}^a_i$ into $-\hbar\delta/\delta A_a^i$ and acting them on the wavefunctional $\Psi(A)$. We have:
\begin{equation}
\hbar{\cal D}_a\frac{\delta \Psi(A)}{\delta A_a^i}=0
\end{equation}
\begin{equation}
\hbar F^i_{ab}\frac{\delta \Psi(A)}{\delta A_a^i}=0
\end{equation}
\begin{equation}
\hbar^2\epsilon^{ij}_k F^k_{ab}\frac{\delta^2 \Psi(A)}{\delta A_a^i\delta A_b^j}=0
\end{equation}
Note that we have chosen the specific ordering that the triads act at left.

In order to get the causal interpretation, one should put the definition $\Psi=R\exp(iS/\hbar)$ into these relations. The result is:
\begin{equation}
{\cal D}_a\frac{\delta R(A)}{\delta A_a^i}=0
\label{G-R}
\end{equation}
\begin{equation}
{\cal D}_a\frac{\delta S(A)}{\delta A_a^i}=0
\label{G-S}
\end{equation}
\begin{equation}
F^i_{ab}\frac{\delta R(A)}{\delta A_a^i}=0
\label{C-R}
\end{equation}
\begin{equation}
F^i_{ab}\frac{\delta S(A)}{\delta A_a^i}=0
\label{C-S}
\end{equation}
\begin{equation}
\epsilon^{ij}_k F^k_{ab}\frac{\delta}{\delta A_a^i}\left( R^2\frac{\delta S(A)}{\delta A_b^j}\right ) =0
\label{H-R}
\end{equation}
\begin{equation}
-\epsilon^{ij}_k F^k_{ab}\frac{\delta S(A)}{\delta A_a^i} \frac{\delta S(A)}{\delta A_b^j}+Q=0
\label{H-S}
\end{equation}
in which the \textit{quantum potential} is defined as:
\begin{equation}
Q=-\hbar^2\epsilon^{ij}_k F^k_{ab}\frac{1}{R}\frac{\delta^2 R(A)}{\delta A_a^i\delta A_b^j}
\end{equation}
Equations (\ref{G-R})--(\ref{C-S}) show the gauge and diffeomorphism invariance of the norm and the phase of the wavefunctional. Equation (\ref{H-R}) is the continuity equation, while equation (\ref{H-S}) is the quantum Einstein--Hamilton--Jacobi equation. The quantum effects as it is always the case in causal interpretation, are introduced via the quantum potential. These are the classical equations\cite{ROVELLI} corrected by the quantum potential.

The quantum trajectories would achieved via the guidance relations:
\begin{equation}
\widetilde{E}^a_i=i\frac{\delta S(A)}{\delta A_a^i}
\end{equation}
\section{Constraints algebra}
In this section we shall study the quantum version of the constraints algebra. In terms of the smeared out Gauss, vector, and scalar constraints:
\begin{equation}
{\cal G}(\Lambda_i)=-i\int d^3x \Lambda_i {\cal D}_a \widetilde{E}^a_i
\label{gauge}
\end{equation}
\begin{equation}
{\cal C}(\vec{N})=i\int d^3x N^b \widetilde{E}^a_i F^i_{ab} -{\cal G}(N^aA^i_a)
\label{diff}
\end{equation}
\begin{equation}
{\cal H}(\rlap{\lower1ex\hbox{$\sim$}}N)=\frac{1}{2}\int d^3x \rlap{\lower1ex\hbox{$\sim$}}N \epsilon^{ij}_k \widetilde{E}^a_i \widetilde{E}^b_j {\cal F}^k_{ab}
\label{time}
\end{equation}
the classical algebra is given by:
\begin{equation}
\{{\cal G}(\Lambda_i),{\cal G}(\Theta_j)\}={\cal G}(\epsilon^i_{jk}\Lambda^j\Theta^k)
\label{gg}
\end{equation}
\begin{equation}
\{{\cal C}(\vec{N}),{\cal C}(\vec{M})\}={\cal C}({\cal L}_{\vec{M}}\vec{N})
\label{dd}
\end{equation}
\begin{equation}
\{{\cal C}(\vec{N}),{\cal G}(\Lambda_i)\}={\cal G}({\cal L}_{\vec{N}}\Lambda_i)
\label{dg}
\end{equation}
\begin{equation}
\{{\cal C}(\vec{N}),{\cal H}(\rlap{\lower1ex\hbox{$\sim$}}M)\}={\cal H}({\cal L}_{\vec{N}}\rlap{\lower1ex\hbox{$\sim$}}M)
\label{dt}
\end{equation}
\begin{equation}
\{{\cal G}(\Lambda_i),{\cal H}(\rlap{\lower1ex\hbox{$\sim$}}N)\}=0
\label{gt}
\end{equation}
\begin{equation}
\{{\cal H}(\rlap{\lower1ex\hbox{$\sim$}}N),{\cal H}(\rlap{\lower1ex\hbox{$\sim$}}M)\}={\cal C}(\vec{K})+{\cal G}(K^aA_a^i)
\label{tt}
\end{equation}
where $K^a=\widetilde{E}^a_i\widetilde{E}^{bi}(\rlap{\lower1ex\hbox{$\sim$}}N\partial_b\rlap{\lower1ex\hbox{$\sim$}}M-\rlap{\lower1ex\hbox{$\sim$}}M\partial_b\rlap{\lower1ex\hbox{$\sim$}}N)$.

In the previous section we saw that the quantum Hamilton--Jacobi--Einstein equation is just the classical one added with the quantum potential. So the quantum trajectories can be obtained from the quantum Hamiltonian given by:
\begin{equation}
H_Q=H+Q
\end{equation}
The smeared out gauge and diffeomorphism constraints would not change but the Hamiltonian constraint is now given by:
\begin{equation}
{\cal H}_Q(\rlap{\lower1ex\hbox{$\sim$}}N)=\frac{1}{2}\int d^3x \rlap{\lower1ex\hbox{$\sim$}}N \epsilon^{ij}_k \widetilde{E}^a_i \widetilde{E}^b_j {\cal F}^k_{ab}+{\cal Q}(\rlap{\lower1ex\hbox{$\sim$}}N)
\end{equation}
where ${\cal Q}(\rlap{\lower1ex\hbox{$\sim$}}N)=\int d^3x \rlap{\lower1ex\hbox{$\sim$}}N Q$. The constraint Poisson bracket (\ref{gg}), (\ref{dd}) and (\ref{dg}) would not change. The Poisson bracket (\ref{dt}) is still valid for the quantum Hamiltonian, because the quantum potential is a scalar density. So we have:
\begin{equation}
\{{\cal C}(\vec{N}),{\cal H}_Q(\rlap{\lower1ex\hbox{$\sim$}}M)\}={\cal H}_Q({\cal L}_{\vec{N}}\rlap{\lower1ex\hbox{$\sim$}}M)
\label{qdt}
\end{equation}
The same is true for the Poisson bracket (\ref{gt}) as the quantum potential is gauge invariant:
\begin{equation}
\{{\cal G}(\Lambda_i),{\cal H}_Q(\rlap{\lower1ex\hbox{$\sim$}}N)\}=0
\label{qgt}
\end{equation}
The difference comes in evaluation of the quantum version of Poisson bracket (\ref{tt}). We have:
\[
\{{\cal H}_Q(\rlap{\lower1ex\hbox{$\sim$}}N),{\cal H}_Q(\rlap{\lower1ex\hbox{$\sim$}}M)\}=
\]
\[
\{{\cal H}(\rlap{\lower1ex\hbox{$\sim$}}N),{\cal H}(\rlap{\lower1ex\hbox{$\sim$}}M)\}+\{{\cal Q}(\rlap{\lower1ex\hbox{$\sim$}}N),{\cal H}(\rlap{\lower1ex\hbox{$\sim$}}M)\}+
\]
\begin{equation}
\{{\cal H}(\rlap{\lower1ex\hbox{$\sim$}}N),{\cal Q}(\rlap{\lower1ex\hbox{$\sim$}}M)\}+\{{\cal Q}(\rlap{\lower1ex\hbox{$\sim$}}N),{\cal Q}(\rlap{\lower1ex\hbox{$\sim$}}M)\}
\end{equation}
the fourth term is zero identically, since the quantum potential is a functional of the connection only. The sum of the second and the third terms is:
\[
\{{\cal Q}(\rlap{\lower1ex\hbox{$\sim$}}N),{\cal H}(\rlap{\lower1ex\hbox{$\sim$}}M)\}+\{{\cal H}(\rlap{\lower1ex\hbox{$\sim$}}N),{\cal Q}(\rlap{\lower1ex\hbox{$\sim$}}M)\}\sim
\]
\[
 i\int d^3z \left ( \rlap{\lower1ex\hbox{$\sim$}}N\epsilon^{ij}_{\ \ k} F^k_{ab}\widetilde{E}^b_j {\cal D}_c(\rlap{\lower1ex\hbox{$\sim$}}M\epsilon^{lm}_i \widetilde{E}^a_l\widetilde{E}^c_m) - \right .
\]
\begin{equation} 
\left . \rlap{\lower1ex\hbox{$\sim$}}M\epsilon^{ij}_{\ \ k} F^k_{ab}\widetilde{E}^b_j {\cal D}_c(\rlap{\lower1ex\hbox{$\sim$}}N\epsilon^{lm}_{\ \ i} \widetilde{E}^a_l\widetilde{E}^c_m) \right )
\end{equation}
where we have used the symbol $\sim$ in order to show that this equality is valid weakly. That is the equation of motion (functional derivative of the Hamiltonian constraint with respect to the connection) is used in its evaluation. A simple calculation then shows that the Poisson bracket of the quantum Hamiltonian with itself is given by:
\begin{equation}
\{{\cal H}_Q(\rlap{\lower1ex\hbox{$\sim$}}N),{\cal H}_Q(\rlap{\lower1ex\hbox{$\sim$}}M)\}\sim 0
\label{qtt}
\end{equation}
which is a result very similar to the one in terms of the old variables\cite{CONSTRAINT}.

At this end it may be useful that obtain the quantum equations of motion by using the Hamilton equations. We have:
\begin{equation}
\dot{A}^i_a=-i\epsilon^{ijk}\rlap{\lower1ex\hbox{$\sim$}}N\widetilde{E}^b_j F_{abk}-N^bF^i_{ab}
\end{equation}
\[
\dot{\widetilde{E}}^a_i=
i\epsilon_i^{jk}{\cal D}_b(\rlap{\lower1ex\hbox{$\sim$}}N\widetilde{E}^a_j\widetilde{E}^b_k) 
\]
\begin{equation}
-2{\cal D}_b(N^{[a}\widetilde{E}^{b]}_i) +\frac{i}{2}\int d^3z \frac{\delta{\cal Q}(\rlap{\lower1ex\hbox{$\sim$}}N)}{\delta A^i_a(z)}
\end{equation}
Also to recover the real quantum general relativity one must set the reality conditions. These are:
\begin{equation}
\widetilde{E}^a_i\widetilde{E}^{bi} \textit{    must be real}
\end{equation}
\begin{equation}
i\epsilon^{ijk}\widetilde{E}^{(a}_i{\cal D}_c\left (\widetilde{E}^{b)}_k \widetilde{E}^c_j\right ) +\frac{i}{2}\int d^3z \frac{\delta{\cal Q}}{\delta A^i_{(a}(z)} \widetilde{E}^{b)i}(x)
\end{equation}
\[
\textit{     must be real}
\]
\section{Conclusion}
We saw that one can successfully construct a causal version of canonical quantum gravity in terms of new variables using the de-Broglie--Bohm interpretation of quantum mechanics. As it is usuall in this theory, all the quantum behavior is coded in the \textit{quantum potential}. Since the theory is a constrained one, one should calculate the constraints algebra and check it for consistency. As in the de-Broglie--Bohm theory any quantum system has a well defined trajectory in the configuration space and one has no operator, the algebra action is in fact the Poisson bracket. We have shown that only the Poisson bracket of Hamiltonian with itself would change with respect to the classical algebra. This Poisson bracket is \textit{weakly}, that is on the equations of motion, equal to zero, different from the classical case where it is \textit{strongly} equal to sum of a gauge transformation and a 3--diffeomorphism. The result is just like to that of the theory in terms of old variables\cite{CONSTRAINT}. This enables one to give the meaning of time generator to the Hamiltonian constraint. At the end the equations of motion are written out and as it is expected the quantum force appears in them.

It must be noted here that all the above results are formal, that is to say, in evaluation of the Poisson brackets and other things we have not regularized the ill--defined terms. For having a rigour result one should evaluate these using a regulator. Introduction of a regulator in general needs to use a background metric and one must show at the end that the result is independent of that background metric. We shall do this in a forthcoming paper.

\end{multicols}
\end{document}